\documentclass[twocolumn,showpacs,superscriptaddress,prb]{revtex4}
\usepackage{amssymb}
\usepackage{amsmath}
\usepackage{graphicx}

\begin{document}


\title{Possible $\pi$-phase shift at interface of two pnictides with antiphase s-wave pairing}

\author{Wei-Qiang Chen and Fu-Chun Zhang}
\affiliation{Department of Physics and Center of Theoretical and
Computational Physics, The University of Hong Kong, Hong Kong,
China}

\date{\today }

\begin{abstract}
We examine the nature of Josephson junction between two identical
Fe-pnictides with anti-phase s-wave pairing. $\pi$-phase shift is found
if the junction barrier is thick and the
two Fe-pnictides are oriented in certain directions relative to
the interface. Our theory provides a possible explanation for the
observed half integer flux quantum transitions in a
niobium/polycrystal NdFeAsO loop, and attributes the $\pi$-phase shift to
intergrain junctions of Fe-pnictides.
\end{abstract}


\maketitle

Since the discovery of a new class of unconventional
superconductors based on $Fe$ compounds, their pairing symmetry
has been one of the most interesting issues. Different from the
high Tc cuprates, where the pairing symmetry is universally
d-wave, there are experimental evidences that pairing symmetry in
the iron pnictides may not be universal. The superconductors with
higher transition temperature $T_c$ are supported by spin singlet
s-wave \cite{chen_bcs-like_2008,
ding_observation_2008,zhang_observation_2008,Tsuei}, while LaFePO
with lower $T_c$ seems to have nodal in its gap
function~\cite{nodal}. It is interesting to note that the iron
pnictide has both hole and electron Fermi pockets, and the
predicted s-wave pairing state has superconducting order
parameters with opposite signs on the electron and hole pockets,
often called anti-phase s-wave or $s_{\pm}$-wave state
\cite{mazin_unconventional_2008, kuroki_unconventional_2008,
Tesanovic, seo_pairing_2008, wang_functional_2009,
chen_strong_2009}. Among the experiments in support of the
$s_{\pm}$ pairing state, the phase sensitive experiment reported
by Chen et al. \cite{Tsuei} provides the most convincing evidence,
where they observed integer and half integer flux quantum
transitions in a niobium/polycrystal NdFeAsO loop. The observed
half integer flux quantum demonstrates the existence of
$\pi$-junctions in the loop of niobium and polycrystal pnictides,
hence a direct evidence for the opposite signs of the
superconducting order parameters in different Fermi pockets. In
passing, we recall that phase sensitive experiments provided a
direct evidence for the $d_{x^2-y^2}$ pairing in superconducting
cuprates\cite{kirtley_symmetry_1995, wollman_experimental_1993}.

Because of the polycrystal nature in the sample of the NdFeAsO, a
phase shift of $\pi$ in the composite loop in tunneling could
occur at the Nb-Nd-1111 interface, or at the junction between two
Nd-1111 grains\cite{Tsuei}.   Theoretically, there have been
several studies to examine the possible $\pi$-phase shift
involving an interface between a conventional superconductor and a
Fe-pnictide with $s_{\pm}$ pairing under certain
conditions~\cite{mazin-david,wqchen2,jphu}, which may help
understand the possible $\pi$-junction at the Nb-Nd-1111
interface. The possibility of a $\pi$-phase shift in the interface
of two Fe-pnictide intergrains of the same doping has not been
carefully examined, although intuitively one may consider it
unlikely. In this paper, we study a Josephson junction between two
iron pnictides.   When the junction barrier is thick and the two
Fe-pnictide grains are oriented in certain directions relative to
the interface, the junction could give a $\pi$-phase shift.

The $\pi$-junction is a Josephson junction with negative critical current,
 and the critical current for junction
between conventional superconductors is defined as
\begin{equation}
\label{eq:1}
  I_{c}\propto \int d\mathbf{k}d\mathbf{q}~\frac{|T|^{2}\Delta _{1}(\mathbf{k}%
    )\Delta _{2}(\mathbf{q})}{E_{1}(\mathbf{k})E_{2}(\mathbf{q})[E_{1}(\mathbf{k}%
    )+E_{2}(\mathbf{q})]},
\end{equation}%
where $T_{\mathbf{k}\mathbf{q}}$ is the tunneling matrix, $E_{i}(\mathbf{k})=%
\sqrt{\epsilon _{i}(\mathbf{k})^{2}+\Delta _{i}(\mathbf{k})^{2}}$ is the quasiparticle energy of the superconductor
$i=1,2$ respectively, $\epsilon _{i}(\mathbf{k})$ is the single electron energy measured relative to the chemical
potential, and $\Delta _{1(2)}$ are superconducting gap functions, which we shall assume to be real here. The
co-efficient of the integral will always be taken to be positive throughout this paper. Eq. ~\eqref{eq:1} can be easily
generalized to the iron pnictide superconductors with multi-bands, and the critical current are given by,
\begin{equation}
\label{eq:2}
I_{c}\propto \sum_{\alpha \beta }\int d\mathbf{k}d\mathbf{q}~\frac{|T_{%
\mathbf{k}\mathbf{q}}^{\alpha \beta }|^{2}\Delta _{1}^{\alpha }(\mathbf{k}%
)\Delta _{2}^{\beta }(\mathbf{q})}{E_{1}^{\alpha }(\mathbf{k})E_{2}^{\beta }(%
\mathbf{q})[E_{1}^{\alpha }(\mathbf{k})+E_{2}^{\beta }(\mathbf{q})]},
\end{equation}%
where $\alpha $ and $\beta $ are index of bands, and all other notations are the same as in the single band expression
except for the additional band indices.  As pointed out by Sigrist and Rice \cite{sigrist}, though the critical current
is not gauge invariant, the parity of the number of the $\pi $-junctions in a loop is gauge invariant.  So in the
following, we choose a convenient gauge where the gap functions of the hole pockets are positive and the gap functions
of the electron pockets are negative.

In the usual case, the junction between two identical
superconductors is a $0$-junction with $I_c$ to be positive. Let
us consider a point junction of two identical pnictide
superconductors.  In this case, $T_{%
\mathbf{k}\mathbf{q}}^{\alpha \beta }= t_0$, independent of
crystal momentum and the band. The critical current is found to be
\begin{equation}
  I_{c}\propto \sum_{\alpha \beta }\int d\mathbf{k}d\mathbf{q}~\frac{\Delta _{1}^{\alpha }(\mathbf{k}%
    )\Delta _{2}^{\beta }(\mathbf{q})}{E_{1}^{\alpha }(\mathbf{k})E_{2}^{\beta }(%
    \mathbf{q})[E_{1}^{\alpha }(\mathbf{k})+E_{2}^{\beta }(\mathbf{q})]}.
\end{equation}%
According to Ambegaokar and Baratoff, the above formula can be further written as
\begin{align}
\label{eq:3}
I_c \propto \sum_{\alpha \beta} \text{sgn}(\Delta_{\alpha}\Delta_{\beta}) N_{\alpha} N_{\beta} \Delta_m K(\sqrt{1 - \frac{\Delta_{m}^2}{\Delta_M^2}}),
\end{align}
where $N_{\alpha}$ is the density of states of Fermi pocket
$\alpha$, $\Delta_m = \min(|\Delta^{\alpha}|, |\Delta^{\beta}|)$
and $\Delta_M = \max(|\Delta^{\alpha}|, |\Delta^{\beta}|)$ are the
smaller and larger gap on the two pockets respectively, and $K$ is
the elliptical integral. In the special case, all the gap
functions have the same amplitude, the elliptical function
$K(0)=\pi/2$, so that $I_c \propto (\sum_{\alpha}
\text{sgn}(\Delta_{\alpha}) N_{\alpha})^2 > 0$.  This positive
definiteness appears to remain valid when the gap amplitudes are
different.  In particular we consider below iron based
superconductor in the 2-dimension limit. We shall work in extended
Brillouin zone for convenience and set the Fe-Fe distance to be
length unit. There are two electron Fermi pockets with one around
$(\pi, 0)$ and one around $(0,\pi)$, and two hole Fermi pockets
around $(0,0)$ point. We consider the case, where the
superconducting gaps on one of the hole pockets and on the
electron pockets are about the same while the gap on another hole
pocket is smaller, as reported in ARPES for
BaFe$_2$As$_2$\cite{ding}. By using the properties of the
elliptical function, $\sqrt{1 - k^2} K(k) \leq K(0) =
\frac{\pi}{2}$, the critical current $I_c$ is found to be always
positive.  This illustrates that the different signs of the
$s-$wave gap functions is necessary condition, but not sufficient
condition for a $\pi$-junction. Below we shall examine a thick
barrier junction between two pnictide superconductors of certain
orientation and show that such a junction may give a $\pi$-phase
shift.

\begin{figure}[htbp]
\centerline{\includegraphics[width=0.4 \textwidth]{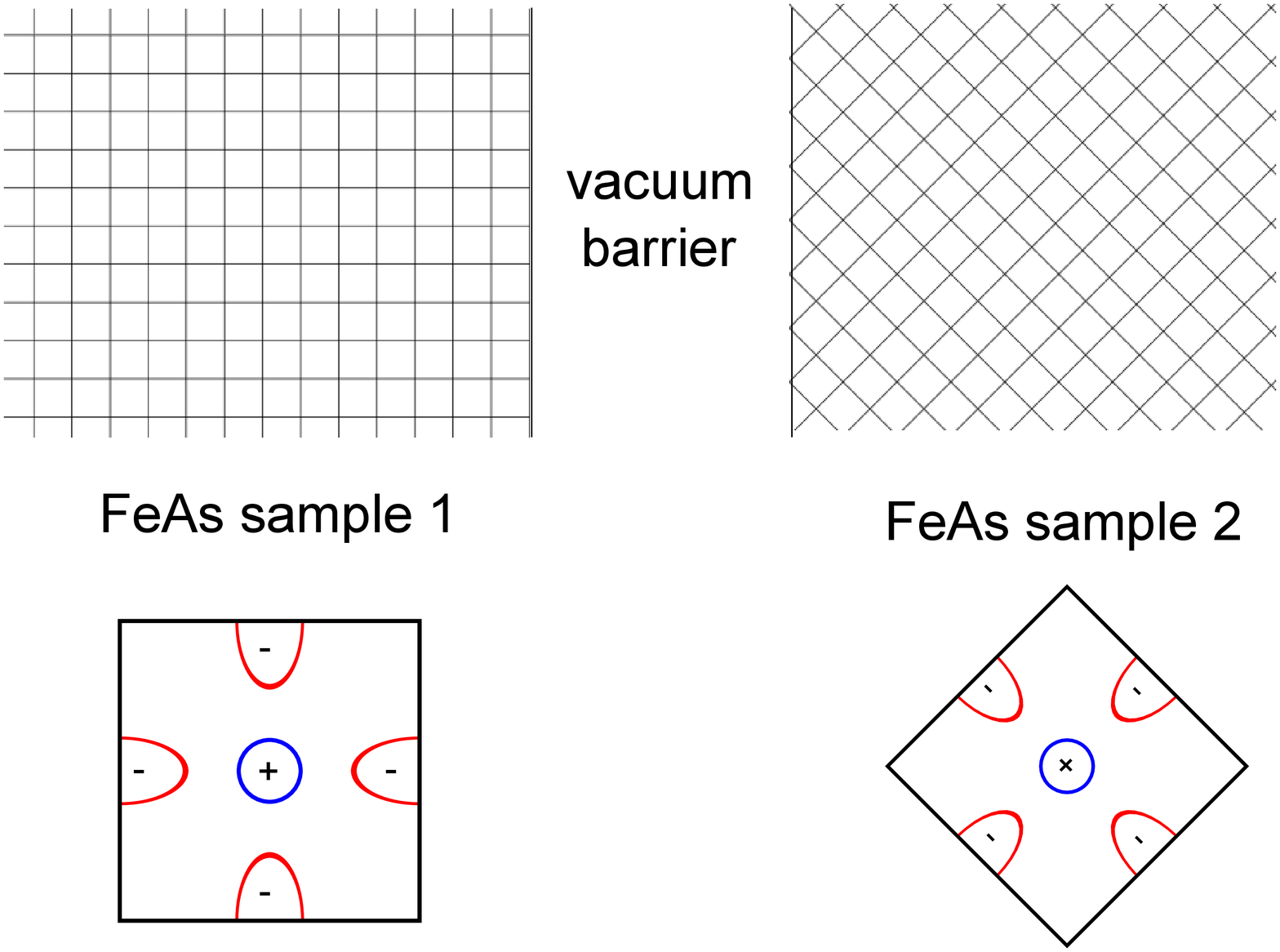}}
\caption{Schematic diagram of a thick Josephson junction of two
iron pnictides separated by a vacuum.  The Fe lattice
  orientation for the two materials are depicted in the upper panel.  Lower panel is the corresponding Fermi surface in
  the Brillouin zone for the two material. }
\label{fig:sketch}
\end{figure}

We consider two half-infinite iron pnictide samples separated by a
vacuum barrier with height $U$ and
width $d$ as shown in fig.~%
\ref{fig:sketch}. The interface is along $yz$ plane so that the
momentum components along $y$ and $z$ directions are conserved in
the tunneling process.  For a thick barrier, the usual assumption
that the tunneling matrix element $T$ is independent of momentum
or crystal momentum is no longer valid.  This can be illustrated
by examining the free electron tunneling process in quasi
one-dimension, where $|T|^{2}$ is the transmission coefficient of
the scattering problem.  For a potential barrier normal to $x$
direction with height $U$ and length $d$, the transmission
coefficient reads
\begin{align}
\label{eq:4}
|T|^{2} & =  \frac{4 \kappa^2k_x q_x}{\kappa^2(k_x +q_x)^2 + (\kappa^2 + k_x^2)(\kappa^2 + q_x^2) \sinh^2(\kappa d) },
\end{align}
where $k_x$ and $q_x$ are the $x$-direction wavevector of the
incoming and outgoing plane waves respectively, $\kappa$ is the
imaginary wavevector inside the barrier with $\kappa =
\frac{1}{\hbar} \sqrt{2m (U - E) + \hbar^2 k_{\parallel}^2 }$, $m$
is the mass of the electron, and $k_{\parallel}$ is the wavevector
parallel to the barrier interface, which is conserved in the
scattering process. In the thick barrier limit, i.e. $\kappa d \gg
1$, one have
\begin{align}
\label{eq:5}
|T|^{2} & \simeq \frac{16 \kappa^2k_x q_x}{(\kappa^2 + k_x^2)(\kappa^2 + q_x^2) } e^{-2 \kappa d}.
\end{align}
If we assume that $p \equiv \frac{1}{\hbar}\sqrt{2m (U - E)} \gg k_{\parallel}$, it can be further simplified as
\begin{align}
\label{eq:6}
|T|^{2} & \propto \frac{\kappa^2k_x q_x}{(\kappa^2 + k_x^2)(\kappa^2 + q_x^2) } e^{-\frac{k_{\parallel}^2}{p} d},
\end{align}
so the transmission coefficient decays exponentially with the
increment of planar wave-vector $k_{\parallel}$. The above formula
can be extended to the electron tunneling process in a lattice
with the following modifications as pointed out by
Mazin\cite{mazin}.
All the wave-vectors in the expression of the above equation are
replaced by the corresponding group velocities except
$k_{\parallel}$ in the exponential factors which tracks the
oscillation of the wavefunction parallel to the interface
direction.  The second modification is to replace the plane
wavefunction in the free electron case by a Bloch wave $\psi_{n
\mathbf{k}}(\mathbf{r}) = e^{i \mathbf{k} \cdot \mathbf{r}}
\omega_{n
  \mathbf{k}}(\mathbf{r})$.
  The periodical function $\omega_{n \mathbf{k}}$ can be
  further Fourier's transformed as
$\omega_{n\mathbf{k}}(\mathbf{r})= \sum_{\mathbf{K}}
F_{n \mathbf{k}, \mathbf{K}} e^{i \mathbf{K} \cdot \mathbf{r}} $,
where $\mathbf{K}$ is reciprocal lattice vector.
 If $\omega_{n\mathbf{k}}(\mathbf{r})$ is localized,
 one can approximate $F_{n \mathbf{k}, \mathbf{K}} \sim$
 constant for not very large $\mathbf{K}$. The wavefunction
can finally be written as
\begin{align}
\label{eq:7}
\psi_{n \mathbf{k}}(\mathbf{r}) & = F_{n \mathbf{k}}
\sum_{\mathbf{K}} e^{i \mathbf{(k+K)} \cdot \mathbf{r}}.
\end{align}
According to eqn.~\eqref{eq:6}, the tunneling is mainly from the component
with $K_{\parallel} = 0$, so the tunneling
matrix reads
\begin{align}
  |T_{\mathbf{k}\mathbf{q}}^{\alpha \beta }|^{2}&\propto A_{\alpha \mathbf{k}, \beta \mathbf{q}} \sum_{\mathbf{K},
    \mathbf{Q}}4m^{2}\hbar ^{2}\kappa^{2} \delta_{\mathbf{(k+K)}_{\parallel }, \mathbf{(q+Q)}_{\parallel }}
  \nonumber\\
  & \phantom{\propto} \times \frac{v_{\alpha\mathbf{k} x} v_{\beta\mathbf{q} x}}{(\hbar ^{2}\kappa^{2}+m^{2}v_{\alpha
      \mathbf{k} x}^{2})(\hbar ^{2}\kappa^{2}+m^{2} v_{\beta \mathbf{q}x}^2)} e^{-\frac{(\mathbf{k+K})_{\parallel}^2
    }{p}d},
\end{align}
where $\mathbf{K}$ and $\mathbf{Q}$ are the reciprocal lattice vector of FeAs
sample 1 and 2 respectively, $\kappa =
\frac{1}{\hbar} \sqrt{2m(U - E) + \hbar^2 (\mathbf{k}+\mathbf{K})_{\parallel}^2 }$,
the delta function tracks the planar
momentum conservation, $A$ is a factor related to the detail information
of the electron wavefunction and can usually be
approximated as constant, and $v_{\alpha \mathbf{k}x}$ and $v_{\beta \mathbf{q}x}$
are the group velocity along $x$
direction of the electron in band $\alpha $($\beta$)
with lattice wavevector $\mathbf{k}$($\mathbf{q})$ respectively.

The exponential factor in above equation can be rewritten as
$e^{-\frac{\mathbf{k}_{\parallel}^2}{p}d}e^{-\frac{(\mathbf{2k+K})_{\parallel}\cdot \mathbf{K}_{\parallel}}{p}d}$, where
the first term is the contribution of the Brillouin zones with $K_{\parallel} = 0$ and the second term is the additional
factor for the Brillouin Zones with finite $K_{\parallel}$.  In the following, we ignore the z-axis dispersion ,which is
very weak in 1111 compound, and only consider the physics of FeAs plane.  For the configuration shown in
fig.~\ref{fig:sketch}, we have that $K_y = 2 n \pi / a$ for FeAs sample 1 and $Q_y = \sqrt{2} m \pi / a$ for FeAs sample
2, where $a$ is the lattice constant.  For a barrier with $d \sim 10a$ and $p \equiv \frac{1}{\hbar}\sqrt{2m (U - E)}
\approx 10 \pi / a$, the largest factor for Brillouin zone with finite $K_{\parallel}$ is $e^{- 2\pi} \approx 0.002$.
So only the contribution from the Brillouin Zone with $K_{\parallel} = 0$ is important.  And because $p^2 \gg
k_{\parallel}^2$, eqn.~\eqref{eq:6} becomes
\begin{equation}
\label{eq:8}
  |T^{\alpha \beta}_{\mathbf{k} \mathbf{q}}|^2 \propto \delta_{{k}_y, {q}_y} v_{\alpha\mathbf{k} x} v_{\beta \mathbf{q}
    x} e^{- \frac{k_{\parallel}^2}{p} d}.
\end{equation}%

Because of the fast drop of the coherence factor $\frac{\Delta}{2E}$ in the critical current when the state is away from
the Fermi surface, the planar momentum conservation parallel to the interface, and the exponential factor in tunneling
matrix in eqn.~\eqref{eq:8}, only the electron pockets around $X$ and the hole pockets of iron pnictide sample 1 and the
hole pockets of sample 2 are important in the configuration shown in Fig. \ref{fig:sketch}.  Because
$\mathbf{k}_{\parallel} \sim 0$, the exponential factor in eqn.~\eqref{eq:8} becomes a constant.  For simplicity, we
assume that the gap $\Delta$ is momentum independent which is consistent with the ARPES result
\cite{ding_observation_2008}.  By substituting eqn.~\eqref{eq:8} into eqn.~\eqref{eq:2} and noticing that velocity $v_x
\propto \frac{\partial \epsilon}{%
\partial k_x}$, the critical current reads
\begin{align} \label{eq:9} I_c &\propto \sum_{\alpha \beta} \int d k_y
  \int^{\epsilon^{\alpha}_{M}(k_y)}_{\epsilon^{\alpha}_{m}(k_y)} d \epsilon_1
  \int^{\epsilon^{\beta}_{M}(k_y)}_{\epsilon^{\beta}_{m}(k_y)} d \epsilon_2 ~
  \frac{\Delta^{\alpha}_1\Delta^{\beta}_2}{E^{\alpha}_1 E^{\beta}_2 [E^{\alpha}_1 + E^{\beta}_2] },
\end{align}
where $\epsilon_m^{\alpha}(k_y)$ and $\epsilon_M^{\alpha}(k_y)$ are the minimum and maximum energy of the electron in
band $\alpha$ with given $k_y$ and any $k_x$. If $\epsilon^{\alpha(\beta)}_m(k_y) - E_f \ll -\Delta_{\alpha(\beta)}$ and
$\epsilon^{\alpha(\beta)}_{M}(k_y) - E_f \gg \Delta_{\alpha(\beta)}$, the integral can be approximated with
$\text{sgn}(\Delta_{\alpha}\Delta_{\beta}) \Delta_m K(%
\sqrt{1 - \Delta^2_m/ \Delta^2_M})$, where $\Delta_m$ and $\Delta_M$ are the smaller one and larger one of amplitude of
the two gaps $\Delta_{\alpha}$ and $\Delta_{\beta}$. If the amplitude of the superconducting gap on those bands are very
close to each other, the elliptical integral $K$ can be approximated as a constant, and the final result is
\begin{align} \label{eq:10} I_c & \propto \sum_{\alpha \beta} \text{sgn}(\Delta_{\alpha}\Delta_{\beta}) \Delta_m
  \min (\lambda_{1\alpha}, \lambda_{2\beta}),
\end{align}
where $\lambda_{i \gamma}$ is the width along the direction parallel to the interface of the Fermi pocket $\gamma$ of
FeAs sample $i$.

\begin{figure}[htbp]
\centerline{\includegraphics[width=0.4\textwidth]{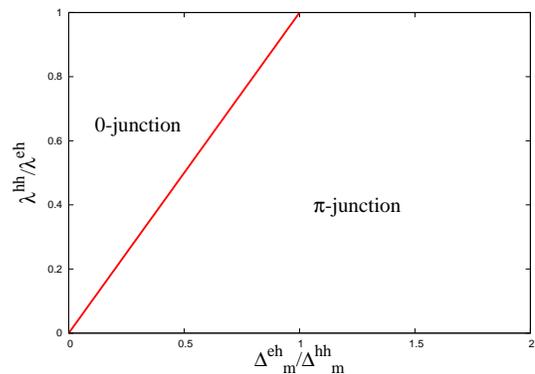}}
\caption{The type of Josephson junction.  The red solid line is $\lambda^{hh}/\lambda^{eh} = \Delta^{eh}_m /
  \Delta^{hh}_m$ which separate the two regions.  See text for details.}
\label{fig:fig2}
\end{figure}

Then we consider the real materials.  Most iron-based
superconductors in 1111 family are electron doped including the
material used in the experiment by Chen et al.\cite{Tsuei}  The
theoretical calculations and experiments show that the parent
compound have two hole pockets around $\Gamma$ point and one
electron pockets around $X$ point. After doping of the electrons,
one would expect the electron pocket expands and the hole pockets
shrink.  when the electron concentration is large enough, the
inner hole pocket will be too small to contribute to the critical
current or even vanished.  In this case, eqn.~\eqref{eq:10}
becomes
\begin{align}
\label{eq:11}
I_c \propto \Delta^{hh}_m \lambda^{hh} - \Delta^{eh}_m \lambda^{eh},
\end{align}
where $\Delta^{\alpha \beta}_m = \min(|\Delta_{1\alpha}|, |\Delta_{2\beta}|)$, $\lambda^{\alpha\beta} =
\min(\lambda_{1\alpha}, \lambda_{2\beta})$.  So the Josephson junction is a $\pi$-junction when $\lambda^{hh}/
\lambda^{eh} < \Delta^{eh}_m / \Delta^{hh}_m$ and vise versa, as shown in fig.~\ref{fig:fig2}.  The ARPES experiment has
shown that the gaps on electron pockets and the outer hole pocket are same \cite{ding_observation_2008},
i.e. $\Delta^{hh}_m = \Delta^{eh}_m$.  So the sign of the critical current is determined by the width of the Fermi
pockets along the direction parallel to the interface.  The DFT result of $LaFeAsO$ has suggest that the hole pocket has
a little anisotropy and the Fermi wavevector along (110) direction is a little larger than the one along (100) direction
\cite{mafj}, i.e. $\lambda^h_1 < \lambda^h_2$ and $\lambda^{hh} = \lambda^h_1$.  On the other hand, if the electron
concentration is large enough, one should have that the width of the electron pocket is larger than the one of the hole
pocket, i.e. $\lambda^e_1 > \lambda^h_2$ and $\lambda^{eh} = \lambda^h_2 > \lambda^{hh} $, which indicates that the
junction is a $\pi$-junction.  Another possibility is that the electron concentration of sample 1 is a littler larger
than the one of sample 2.  In this case, the size of hole pocket of sample 1 should be smaller than the one of sample 2,
i.e. $\lambda^h_1 < \lambda^h_2$.  Again, if the electron concentration is large enough, one should have $\lambda^e_1 >
\lambda^h_2$, and the junction should also be a $\pi$-junction.

For the material in 122 family, both electron and hole can be doped with various ways.  For the electron doped 122
material, the ARPES experiment suggests that there is only one hole pocket around $\Gamma$ point, and the widths of hole
pocket and electron pocket are comparable\cite{ding} which should be similar with the e-doped 1111 case.  So one should
observe similar phenomena in the e-doped 122 material with large enough electron doping.  But the situation is different
in the hole doped case where both of the two hole pockets will contribute to the critical current, and the size of hole
pockets should be larger than the one of electron pocket.  By taking account that the gap on outer hole pocket is same
as the one on electron pocket, one can find that the critical current is positive according to eqn.~\eqref{eq:10}, and
one can not observe half-flux jump in those hole doped 122 material.

In summary, we have examined the Josephson junctions between two
Fe-pnictides with the same electron concentration. We found that
it is impossible for two Fe-pnictides to form a $\pi$-junction in
the case that tunneling matrix element is momentum independent.
But a $\pi$-junction can be formed in the thick barrier case where
the pockets with large parallel momentum are irrelevant.  Such
$\pi$-junctions may be the origin of the $\pi$-flux jump in C.-T.
Chen et al.'s experiment.

We wish to thank fruitful discussions with C.C. Tsuei and C. T.
Chen on their experiments and the related physics. We acknowledge
partial financial support from Hong Kong RGC grants ...


\begin{thebibliography}{9}
\bibitem{chen_bcs-like_2008} T.~Y. Chen {\em et~al.}, Nature {\bf 453}, 1224 (2008).

\bibitem{ding_observation_2008} H.~Ding {\em et~al.}, Euro. Phys. Lett. {\bf 83}, 47001 (2008).

\bibitem{zhang_observation_2008} X.~H. Zhang {\em et~al.}, Phys. Rev. Lett. \textbf{102}, 147002 (2009).

\bibitem{Tsuei} C.-T. Chen, C. C. Tsuei, M. B. Ketchen, Z.-A. Ren, and Z. X. Zhao, Nature Physics {\bf 6},260 (2010).

\bibitem{nodal} Clifford W. Hicks {\em et~al.}, Phys. Rev. Lett. \textbf{103}, 127003 (2009)

\bibitem{mazin_unconventional_2008} I.~I. Mazin {\em et~al.}, Phys. Rev. Lett. {\bf 101}, 057003 (2008).

\bibitem{kuroki_unconventional_2008} K.~Kuroki {\em et~al.}, Phys. Rev. Lett. {\bf 101}, 087004 (2008).

\bibitem{Tesanovic} V. Cvetkovic and Z. Tesanovic, Euro. Phys. Lett. \textbf{85}, 37002 (2009).

\bibitem{seo_pairing_2008} K.~Seo, B.~A. Bernevig, and J.~P. Hu, Phys. Rev. Lett. {\bf 101}, 206404 (2008).

\bibitem{wang_functional_2009} F.~Wang {\em et~al.}, Phys. Rev. Lett. {\bf 102},
  047005 (2009);Zi-Jian Yao, Jian-Xin Li, and Z D Wang, New J. Phys. \textbf{11},025009 (2009).

\bibitem{chen_strong_2009} Wei-Qiang Chen {\em et~al.},
Phys. Rev. Lett. {\bf 102}, 047006 (2009).

\bibitem{kirtley_symmetry_1995} C. C. Tsuei {\em et~al}, Phys. Rev. Lett. 73, 593 (1994).

\bibitem{wollman_experimental_1993} D.~A. Wollman {\em et~al}, Phys. Rev. Lett.  {\bf 71}, 2134 (1993).

\bibitem{mazin-david} David Parker and Igor Mazin, Phys. Rev. Lett. 102, 227007 (2009)

\bibitem{jphu} Wei-Feng Tsai {\em et~al}, Phys. Rev. B \textbf{80}, 012511 (2009).

\bibitem{wqchen2} Wei-Qiang Chen, Fengjie Ma, Zhong-Yi Lu, and Fu-Chun Zhang, Phys. Rev. Lett. \textbf{103}, 207001 (2009)

\bibitem{sigrist} M.~Sigrist and T.~M. Rice, J. Phys. Soc. Jpn. {\bf 61}, 4283 (1992).

\bibitem{mazin} I. I. Mazin, Europhys. Lett. \textbf{55}, 404 (2001)

\bibitem{ding} K. Terashima et al., PNAS \textbf{106}, 7330 (2009)

\bibitem{mafj} Fengjie Ma, private communication
\end{thebibliography}
\end{document}